# Computational Enhancement of Molecularly Targeted Contrast-Enhanced Ultrasound: Application to Human Breast Tumor Imaging

Andrew A. Berlin, Mon Young, Ahmed El Kaffas, Sam Gambhir,
Amelie Lutz, Maria Luigia Storto, and Juergen Willmann

*Abstract*—Molecularly targeted contrast enhanced ultrasound (mCEUS) is a clinically promising approach for early cancer detection through targeted imaging of VEGFR2 (KDR) receptors. We have developed computational enhancement techniques for mCEUS tailored to address the unique challenges of imaging contrast accumulation in humans. These techniques utilize dynamic analysis to distinguish molecularly bound contrast agent from other contrast-mode signal sources, enabling analysis of contrast agent accumulation to be performed during contrast bolus arrival when the signal due to molecular binding is strongest.

Applied to the 18 human patient examinations of the first-in-human molecular ultrasound breast lesion study, computational enhancement improved the ability to differentiate between pathology-proven lesion and pathology-proven normal tissue in real-world human examination conditions that involved both patient and probe motion, with improvements in contrast ratio between lesion and normal tissue that in most cases exceed an order of magnitude (10x). Notably, computational enhancement eliminated a false positive result in which tissue leakage signal was misinterpreted by radiologists to be contrast agent accumulation.

## I. INTRODUCTION

MOLECULARLY targeted ultrasound contrast agents preferentially bind to a molecular target of interest, accumulating in the body and generating a signal that is observable by imaging techniques such as contrast-mode ultrasound [15]. One of the major challenges when imaging contrast agent accumulation is that the signal from accumulating contrast agent is confounded by signal from other sources. Signal from contrast-mode *tissue leakage*, such as connective tissue that appears to be contrast agent; signal from contrast agent that has accumulated *non-specifically* for reasons other than molecular binding; and signal from contrast agent particles that are *freely flowing* in the bloodstream, are all intermixed with the signal of accumulating contrast agent associated with molecular binding [2,16]. We have employed computational enhancement techniques to de-confound these intermixed signals, enabling imaging during the period of high free-flow of contrast agent in the bloodstream, near the time of contrast bolus arrival, when the molecular signal from specific binding is strongest.

We demonstrate these techniques using data acquired by Willmann et al. [1] during the first-in-human molecularly targeted ultrasound study of suspected human breast cancer lesions, which employed BR-55, a microbubble contrast agent functionalized to molecularly bind to the KDR protein, as an indication of angiogenesis activity. Data analysis in that study was performed manually by skilled radiologists, primarily using data acquired several minutes after arrival of the contrast bolus, after the flow of free contrast agent in the bloodstream

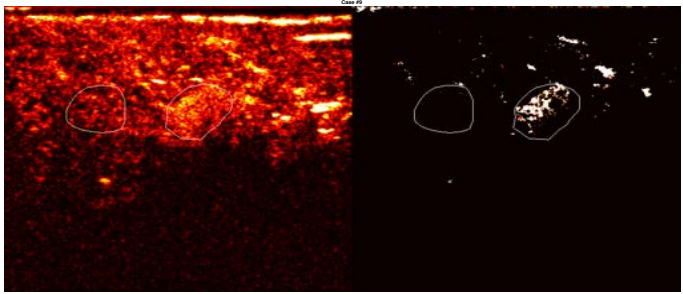

*Figure 1: Results of Computational Enhancement. Left: raw mCEUS image 36 seconds after contrast bolus arrival. Right: Computational enhancement results following tissue leakage and flow disambiguation. The outlined region on the left in both images corresponds to pathology proven normal tissue, while the outlined region on the right corresponds to pathology-proven lesion tissue. (case #9-129)*

Preprint posted 6/21/20.
This work was supported in part by the Canary Foundation, by the Charles Stark Draper Laboratory, and by Bracco Corporation.
Andrew Berlin is with the Charles Stark Draper Laboratory, Cambridge, MA, andrew.berlin@ieee.org.
Mon Young is with the Charles Stark Draper Laboratory, Cambridge, MA, myoung@draper.com.
Juergen Willmann (deceased) was with Stanford Medicine.
Amelie Lutz is with Stanford Medicine, Stanford, CA, alutz@stanford.edu.
Maria Luigia Storto is with Bracco Diagnostics Corporation, MariaLuigia.Storto@diag.bracco.com.
Ahmed El Kaffas is with Stanford Medicine, Stanford, CA, elkaffas@stanford.edu.
Sam Gambhir is with Stanford Medicine, Stanford, CA, sgambhir@stanford.edu.
The authors would like to thank Don Listwin for suggesting this collaboration.

had largely subsided. We show that use of computational enhancement can make it practical to image earlier, during and shortly after contrast bolus arrival when the signal is strongest, before substantial unbinding or bubble destruction occurs. Another advantage to imaging during contrast bolus arrival is that accumulation rate information can be captured to help disambiguate specific binding to the molecular target vs. non-specific accumulation.

Our computational enhancement pipeline explicitly models and removes tissue leakage signal. It also provides a qualitative estimate of the amount of accumulated contrast agent present, even during periods of high flow when the molecularly bound microbubble signal cannot be directly observed because it is obscured by signal from free-flowing contrast agent. Computational enhancement provides substantial improvements in contrast ratio between pathology-proven lesion and pathology-proven non-lesion tissues, as illustrated by the computational enhancement results shown in Figure 1.

## II. IMAGING CHALLENGES

Imaging of contrast agent accumulation in humans poses unique challenges relative to imaging in small laboratory animals. First, the imaging geometries and power levels involved induce significant amounts of tissue leakage signal, typically associated with connective tissue that has similar resonant properties to contrast agent, and hence can evade the harmonic filters employed in commercial ultrasound machines. The most effective approaches to distinguish and eliminate signal due to tissue leakage involve destruction of the microbubble contrast agent using a high energy ultrasound pulse [16]. Any signal that remains after bubble destruction is due to tissue leakage. Due to safety concerns about ultrasound-induced bubble destruction potentially causing damage to small blood vessels [3], destructive protocols were not employed in the human breast imaging study, necessitating development of an alternative, non-destructive method to distinguish accumulating contrast agent from tissue leakage.

When imaging at modest frame rates during periods of high flow of microbubbles in the bloodstream, the signal from specifically-bound accumulating microbubbles is masked by signal from flowing contrast agent in *every* captured image frame. Thus clever techniques for non-destructively disambiguating between flowing and stationary microbubbles, such as the Minimum-Intensity-Projection methods described in [2], which rely on at least occasional absence of flowing contrast agent, are not effective during periods of high flow. To address this issue, we developed a statistical method that *estimates* the degree of accumulation, even during periods of high flow when it cannot be directly observed.

## III. RELATED WORK

### A. Estimation based on Intensity Projection

In situations for which the flow of contrast agent is modest or has partially subsided, several minutes after contrast bolus arrival, we have found that the Minimum Intensity Projection (MinIP) and Percent Intensity Projection (PerIP) techniques described in [2] are highly effective for removing signal associated with flowing contrast agent. These techniques take advantage of the fact that intensity increases associated with flowing contrast agent will only be present in some of the frames, so occasionally a frame without flowing contrast agent will be captured. Taking the minimum intensity value across a multi-frame time window reveals the portion of the intensity that is associated with bound contrast agent, since the presence of flowing contrast agent occurs only occasionally. However, when imaging in the presence of high concentrations of flowing contrast agent, the signal due to accumulation is typically not independently observable in any frame, causing the MinIP and PerIP techniques to lead to false positive indications in which flowing contrast agent is mistaken to be accumulated contrast agent.

### B. Multi-compartment Models

Turco et al.[11] introduce a flow modeling approach that fits a multi-compartment model of the rate of the anticipated flow dynamics and rate of accumulation of contrast agent. This approach assumes a single time of arrival and profile of the contrast bolus, and models accumulation relative to that arrival time. In the Willmann et al. 2017 study data, it is not uncommon for lesions to have multiple feeding blood vessels, as well as vessels that fold back on themselves and appear multiple times in the same imaging plane. This creates an overlapping multiplicity of bolus arrival times that can make it somewhat challenging to fit model parameters. A promising area for future work would be to examine ways to employ multiple overlapping compartment models of the form suggested by Turco et al. to account for these deviations from the idealized single arrival time first-pass model.

### C. Overcoming Contrast-mode Tissue Leakage

A popular alternative to bubble destruction is tissue background subtraction using a reference image acquired prior to contrast bolus arrival. When post-contrast and pre-contrast images are perfectly aligned, subtraction will eliminate the tissue leakage background. However, due to patient motion, organ motion, probe motion, and changes to probe orientation, using a pre-contrast reference frame, or even an average of several pre-contrast frames, as the basis for subtraction throughout a 60-90 second imaging session is only minimally effective. In particular, out-of-plane motion of the probe, which is common, can cause an area of tissue leakage signal that is not present in the pre-contrast reference frame to suddenly appear and be misinterpreted as contrast agent accumulation. Additionally, varying probe pressure can dynamically change the pattern of the tissue leakage signal, creating changes in background signal that can go well beyond simple translation or rotation. This necessitates use of a tissue leakage removal method that accounts for lateral and out-of-plane motion, for tissue compression due to varying applied probe pressure, and for the intermittent nature of signal leakage through harmonic contrast-mode filters.

## IV. COMPUTATIONAL ENHANCEMENT SYSTEM ARCHITECTURE

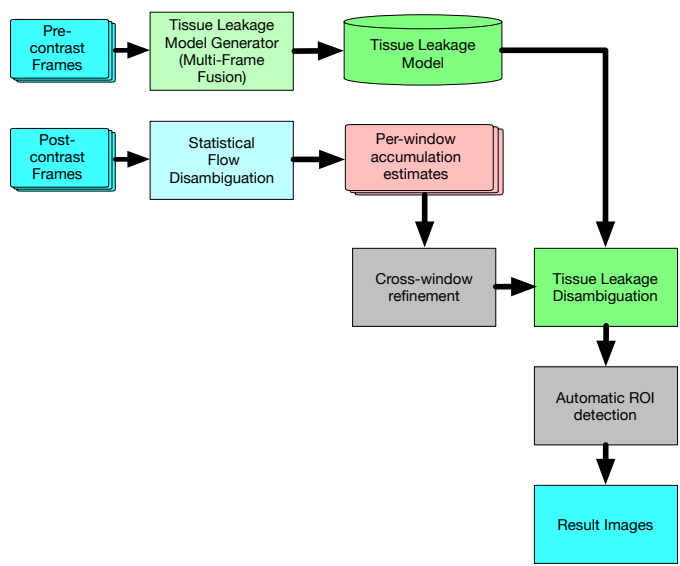

*Figure 2: System Architecture Block Diagram*

Our computational enhancement system architecture, shown in Figure 2, encompasses four distinct computational enhancement modules:
1. *Tissue leakage disambiguation*.
2. Cross-window intensity analysis to distinguish non-specific from specific binding.
3. *Flow disambiguation*.
4. Automatic ROI identification.

This article focuses on the *tissue leakage* and *flow disambiguation* modules, showing results obtained with only those two enhancements activated.

## V. TISSUE LEAKAGE DISAMBIGUATION

### A. Tissue Leakage Modeling via Multi-Frame Fusion

To non-destructively identify tissue leakage signal, we utilize a tissue leakage modeling approach that fuses multiple frames captured over the course of 30-45 seconds (at one frame per second) prior to contrast-agent bolus arrival. This multi-frame fusion forms an estimate of which spatial locations potentially contain tissue leakage signal components. The use of an extended period of time to observe pre-contrast arrival signals permits the effect of patient and operator-induced probe motion to be captured and represented. It also provides ample time for intermittent tissue leakage signals that occasionally evade the ultrasound machine's contrast-mode filters to be detected and to be incorporated into the tissue leakage model.

For early disease screening, to minimize false positives, use of a highly conservative modeling approach is desirable. Any contrast signal that is detected prior to contrast arrival, even if only present momentarily, is considered to be tissue leakage signal throughout the examination, so as to minimize false positive detections. For this purpose, we have found that a Maximum Intensity Projection (MaxIP) across 30 or more pre-contrast frames is quite effective, as illustrated in Figure 5, which shows that the pre-contrast MaxIP approach substantially improves lesion/normal contrast ratio relative to use of a traditional single frame background model. For applications such as tumor margin detection for pre-surgical planning, a less aggressive background model is preferable, erring in favor of classifying a questionable intermittent signal as tumor, to ensure that tumor margins are not underestimated. This paper focuses on the early disease screening scenario, using the maximum intensity projection approach.

The maximum intensity projection interprets contrast signal observed in any frame prior to contrast bolus arrival to be tissue leakage background signal, generating a composite tissue leakage model image (Figure 3). Empirically, across the 18 human breast imaging cases we have studied, this maximum intensity projection approach to background modeling has proven to be an effective approximation of the range of probe and patient motion that is exhibited a few seconds later during the period of contrast bolus arrival.

### B. Quantifying Examination Quality: The Spread Ratio

Probe and patient movement during the pre-contrast arrival period leads to background signal associated with tissue leakage to appear to be laterally shifted in the case of in-plane probe/patient relative motion, or to appear or disappear (for out-of-plane probe/patient motion). When the maximum intensity projection is applied across multiple pre-contrast arrival image frames, image displacement due to probe-patient relative motion leads to enlargement or 'spreading' of tissue leakage background signal intensity, and to persistence of intermittent tissue leakage signal. This 'spreading' may be quantified by taking the ratio of the overall intensity (across all pixels) of the fused background signal model to the overall intensity of a single pre-contrast arrival frame. In the ideal case of absence of patient motion, probe motion, and measurement error, this ratio will be 1. As motion increases, the ratio increases, as the image intensity spreads across multiple pixels of the maximum intensity projection.

The spread ratio can serve as a measure of examination quality that may be used to further validate the confidence level in the diagnostic conclusions drawn from the molecular imaging data, i.e. whether the samples gathered after contrast bolus arrival can be expected to be free of substantial motion artifacts. As illustrated in Figure 4, while most examinations have modest spread ratios, case #5 exhibits substantially more probe-patient relative motion, which is captured in the elevated spread ratio. The spread ratio can be used to provide real-time feedback to the ultrasound operator, indicating whether a sufficient level of probe and patient stability have been achieved to warrant release of the contrast agent bolus.

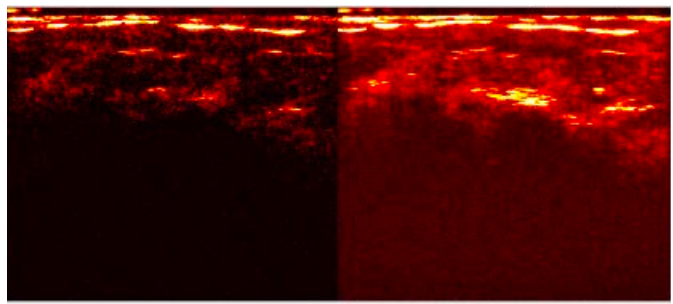

*Figure 3: Tissue leakage signal prior to contrast arrival (case 16). Left: Tissue leakage model derived from a single frame. Right: Composite model derived from multi-frame maximum intensity fusion across 34 pre-contrast arrival frames. Note how the nature of the intermittent tissue leakage signal, captured in the multi-frame fusion, if not removed could easily be misinterpreted as being due to contrast agent, even though there is no contrast agent present.*

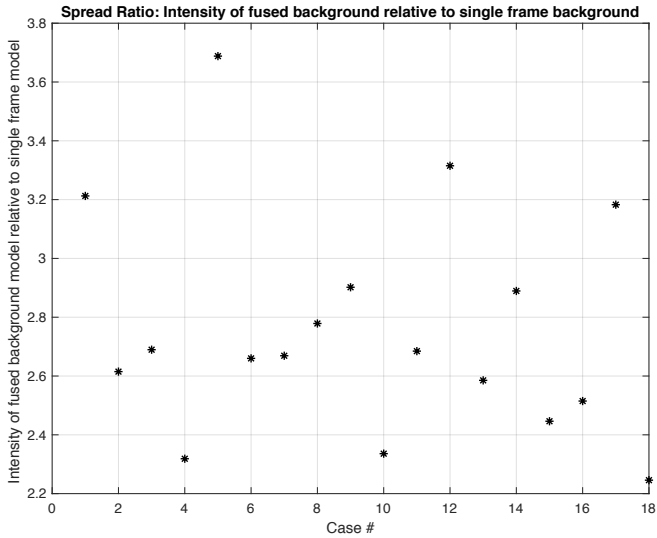

*Figure 4: Spread ratio measuring overall intensity of the multi-frame fusion model relative to overall intensity of a single representative frame prior to contrast arrival. The outliers, such as case 5, are an indication of a case whose examination involved more than typical patient-probe relative motion.*

## VI. FLOWING CONTRAST AGENT DISAMBIGUATION

### A. Advantages of imaging during bolus arrival

Due to molecular binding of bubbles to target molecules, molecular ultrasound offers a unique opportunity to obtain still images post wash-in, after the flow of contrast agent has subsided. However, there are substantial advantages to imaging during wash-in. First, not waiting for the flow to subside dramatically decreases the time elapsed from the construction of the tissue leakage background subtraction model to the time of measurement, thereby minimizing changes in probe and patient positioning during contrast bolus arrival relative to those present prior to bolus arrival. Second, imaging during wash-in permits the binding accumulation rate to be captured, which can help distinguish between specific and non-specific binding. Third, significant unbinding occurs while waiting for the flow of contrast agent to subside, so the signal from bound contrast agent is stronger when imaging soon after molecular binding occurs. Far greater contrast between lesion and normal tissue is available through computationally-enhanced imaging during wash-in rather than waiting several minutes for contrast agent flow to subside.

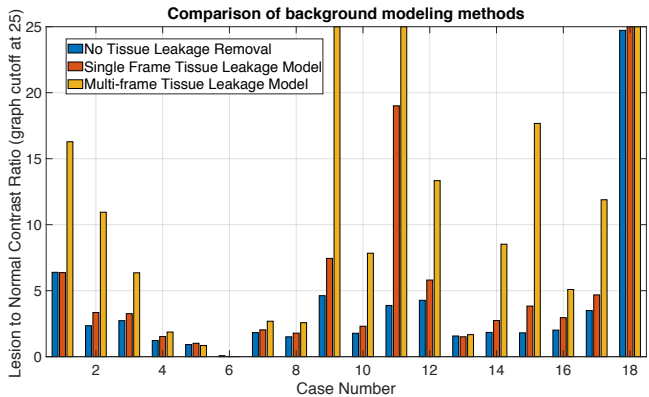

*Figure 5: Comparison of Tissue Leakage Background Signal Modeling Methods. Once flowing contrast agent has been removed, in this case using the Minimum Intensity Projection method described in [2], disambiguation of lesion vs. normal can be further improved through removal of tissue leakage signal using the multi-frame maximum-intensity-projection model.*

Distinguishing stationary contrast agent from flowing contrast agent is extremely difficult for human readers to achieve in the presence of a high concentration of flowing contrast agent, since contrast agent appears throughout the field of view, as shown in Figure 1. Detailed intensity analysis of individual pixels and time points makes it possible to readily estimate macro-scale properties such as contrast bolus arrival time and wash-in duration, but it is not at all clear from the raw data (Figure 6) what portion of the contrast agent intensity is due to molecularly bound contrast agent vs. flowing contrast agent. To overcome flow disambiguation challenges, the radiologists reading the results of the initial human breast imaging study primarily utilized later time points, several minutes after contrast bolus arrival, after the high concentration of flowing contrast agent had subsided.

### B. Flow Disambiguation using Temporal Windowing

Stationary contrast agent can be distinguished from flowing contrast agent using time-based windowing techniques. As illustrated in Figure 7, samples captured by the ultrasound Cineloop are grouped into overlapping time windows. These windows are each used for statistical analysis, within windows to estimate accumulation level and across windows for accumulation rate estimation.

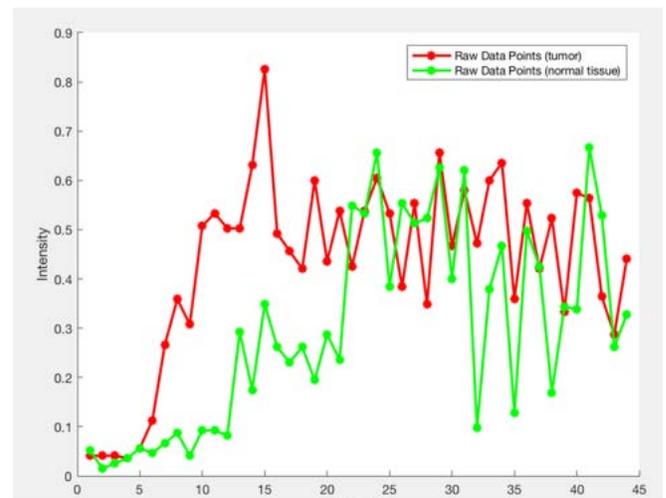

*Figure 6: Examples of unprocessed CEUS data during the first 45 seconds after contrast bolus administration. Red: a single pixel signal from a tumor lesion. Green: A single pixel signal from a normal tissue region. (Case 17/151).*

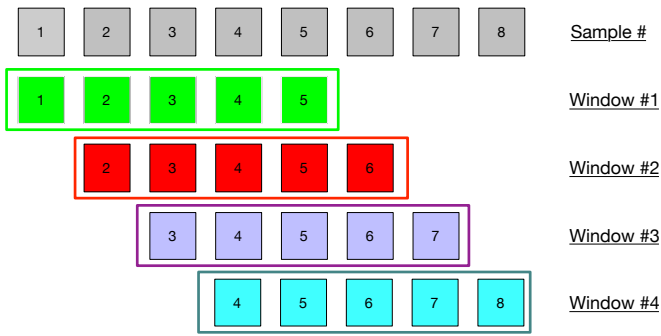

*Figure 7: Measurement windowing approach, grouping adjacent samples to form statistical measurement windows. For example, window #3 includes samples #3,#4,#5,#6, and #7.*

Grouping samples into windows is advantageous because it permits analysis of samples over relatively smaller time scales during which key parameters such as mean signal intensity are more uniform than is the case over larger time periods. For the breast imaging dataset [1], we have found window size W=20 to be effective at a sample rate of 1 sample per second.

### C. Flow disambiguation within a temporal window

In the absence of measurement error or noise, for low to moderate concentrations of non-stationary contrast agent, the minimum intensity projection (MinIP) across the samples in a time window will reflect the portion of the signal in that window that is due to stationary contrast agent particles. When concentration of flowing contrast agent is low, it is likely that the measurement window will include at least one sample for which no flowing contrast agent particles is present at a particular pixel/region. If one waits long enough after contrast agent injection prior to measuring, flowing contrast agent concentration will have decreased sufficiently that there will be several such samples within a measurement window.

The availability of multiple valid samples without flow permits the use of alternative projections that are more resilient to measurement noise, such as the 20% projection (PerIP) approach suggested in [2], that takes the intensity of the weakest 20% of samples to be reflective of the concentration of bound contrast agent. The PerIP approach provides resilience to drop-out measurement noise in which a single time sample has an erroneously low value.

In high flow situations the MinIP approach becomes error prone because there will be flowing contrast agent present in **all** of the acquired samples within each window. The projection no longer reflects the intensity due only to the non-flowing contrast agent signal, but also includes intensity associated with some of the flowing contrast agent. If one uses the PerIP approach, the situation gets even worse, since in high flow, finding a window in which 20% of the samples are not due to flowing contrast agent is nearly impossible. To address this issue, we developed an alternative method that can estimate stationary contrast agent concentration within a window *even when it cannot be observed directly due to presence of flowing contrast agent*.

### D. Statistical estimation approach

The detailed dynamics of blood flow in normal and lesion tissue encountered in the human breast data set are quite complex. Some lesions are fed by more than one blood supply source. Vessels follow tortuous trajectories and may loop back towards the lesion, creating multiple 'arrival times'. There tend to be multiple binding sites present within each pixel, with multiple probabilistic events contributing to the overall observed intensity. Due to the central limit theorem, a combination of random variables tends towards the normal distribution. Thus contrast agent accumulation may be approximated based on aggregate dynamic effects using the statistical mean-offset approach outlined below.

The contribution of contrast agent accumulation to the overall observed signal is estimated based on statistical properties across the entire collection of samples within the measurement time-window, rather than relying on one minimum sample value as the MinIP approach does. The statistical approach models the stationary contrast agent intensity at each location '$s$' in terms of standard deviations below the mean at that location:

$$s = max\ (0, u - \alpha\ \sigma\ )$$
**Equation 1: Statistical estimation (mean-offset) approach**

In Equation 1, $\sigma$ is the standard deviation of the intensity of the samples within the window, $u$ is the mean intensity of the samples within the window, and $\alpha$ is an algorithmic parameter to be selected and tuned. We have found that it is effective to use a single value of $\alpha$ across all 18 cases in the Willmann breast imaging data set [1], rather than adjusting $\alpha$ for each patient. $\alpha = 1.7$ produces images that are visually indistinguishable from those produced by the Minimum Intensity Projection.

Relative to the MinIP approach, the statistical approach with moderate value of $\alpha$ provides greater resilience to single-sample dropout noise and reduced latency response to intensity increases. This provides finer time-resolution information for cross-window estimation of accumulation rate, which can help in disambiguation of specific from non-specific binding.

To qualitatively estimate the intensity level that is due to bound contrast agent, which in high flows is not directly visible via minimum intensity projection, it is effective to increase $\alpha$, which has the effect of reducing the intensity level estimates across the image in a manner that is flow-dependent. For qualitative imaging purposes, for early disease detection $\alpha = 2.7$ has been found to be quite effective across all 18 cases to optimize the contrast between lesion and normal tissue while minimizing flow-based imaging artifacts. We anticipate that in clinical practice, $\alpha$ may be predetermined, or may be tuned manually (for instance the equivalent of a 'contrast' knob on a display) until imaging artifacts disappear from the non-lesion portions of an image. Alternatively, $\alpha$ can be tuned automatically to optimize image quality metrics.

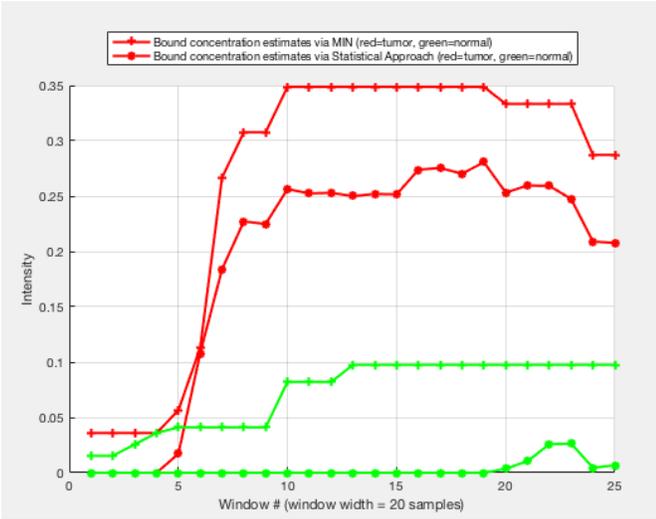

*Figure 8: Statistical estimation of accumulation for the raw data shown in Figure 6. Red = tumor; Green = normal. Vertical tick mark: MinIP accumulation estimate. Circular tick mark: Statistical mean-offset accumulation estimates ($\alpha = 2.5$).*

### E. Advantages of the Statistical Approach

The statistical mean-offset approach is advantageous in part because mean and standard deviation are properties computed from consideration of all samples within the measurement window, reducing sensitivity to a single noisy reading (which could produce a false minimum value). More importantly, this approach does not require availability of a sample in which no flowing contrast agent is present, making it suitable for imaging scenarios that involve high concentrations of flowing contrast agent.

Finally, when window-based analysis results are analyzed across multiple overlapping windows to produce accumulation rate estimates, the statistical approach provides finer time resolution information about contrast agent accumulation levels within each window than the minimum intensity projection does. In the statistical approach, accumulation level updates occur continuously as each new data point is received. In contrast, for the Minimum Intensity Projection approach for a window of size W, a single MINIMUM sample value persists over the course of W measurement windows, masking details of further accumulation. Hence during washin the MinIP approach provides accumulation information as a series of steps occurring every W windows, while the statistical approach provides finer time-resolution information. Availability of finer time resolution information opens the door to development of methods to disambiguate regions of molecularly-specific accumulation of contrast agent from regions of non-specific accumulation (such as blood vessels) based on accumulation rate information obtained via cross-window analysis of the accumulation levels exposed by the statistical approach to computational enhancement.

### F. Morphological Processing

The statistical estimation approach produces sufficiently high contrast ratios between lesion and normal tissue that it becomes practical to apply morphological processing operations to fill-in tiny spatial gaps without causing undo amplification of non-accumulation signals. For the results reported in this article, a morphological closure operation using a disk-shaped structuring element of radius 2-pixels was applied to the flow-removed image prior to tissue leakage removal.

## VII. COMPUTATIONAL ENHANCEMENT RESULTS

Figure 8 shows the results of applying the statistical approach to the raw data example shown in Figure 6. For comparison purposes, the Minimum Intensity Projection results are also shown. The delineation between normal and lesion tissues becomes quite clear. This added clarity is reflected visually in the image-based results in Figure 10 and in the Lesion/Normal contrast ratio improvement factor in Figure 9.

As shown in Figure 9 and Figure 11, in comparison to the approach used in the first-in-human study of waiting for the flow of contrast agent to subside [1], our computational enhancement approach, using a combination of maximum intensity projection tissue-leakage signal removal and statistical estimation-based flow-removal, achieves substantially improved lesion/contrast ratio and eliminates a false positive result (case #6). Additionally, Figure 12 shows a comparison of the improvements made available via the statistical approach in comparision to the MinIP approach, during the high concentration of flowing contrast agent that is present when imaging during contrast bolus arrival.

## VIII. CONCLUSIONS

mCEUS imaging in humans without use of bubble destruction, in the presence of patient motion, probe motion, and organ motion, with imaging geometries that include significant amounts of tissue leakage, poses unique challenges relative to imaging in immobile laboratory animals in which imaging protocols that incorporate bubble destruction can be utilized. We have shown that these challenges can be overcome, making it practical to image contrast accumulation in humans during and shortly after contrast bolus arrival, without having to wait for the flow of contrast agent to subside. This early analysis, while flowing contrast agent is still present in high concentration, permits imaging when the signal is strongest, before substantial unbinding or bubble destruction occurs.

Accounting for tissue leakage by using a background model that accounts for patient, probe, and local tissue motion substantially improves the lesion/normal contrast ratio, helping to prevent intermittent appearance and disappearance of tissue leakage signal from being mistaken for contrast agent accumulation. The use of statistical estimation methods to predict the amount of contrast agent accumulation, rather than relying on occasional direct observation of contrast accumulation, or requiring that the bolus arrival match a pre-envisioned model, permits analysis even at high flow rates without incurring undue false positive results.

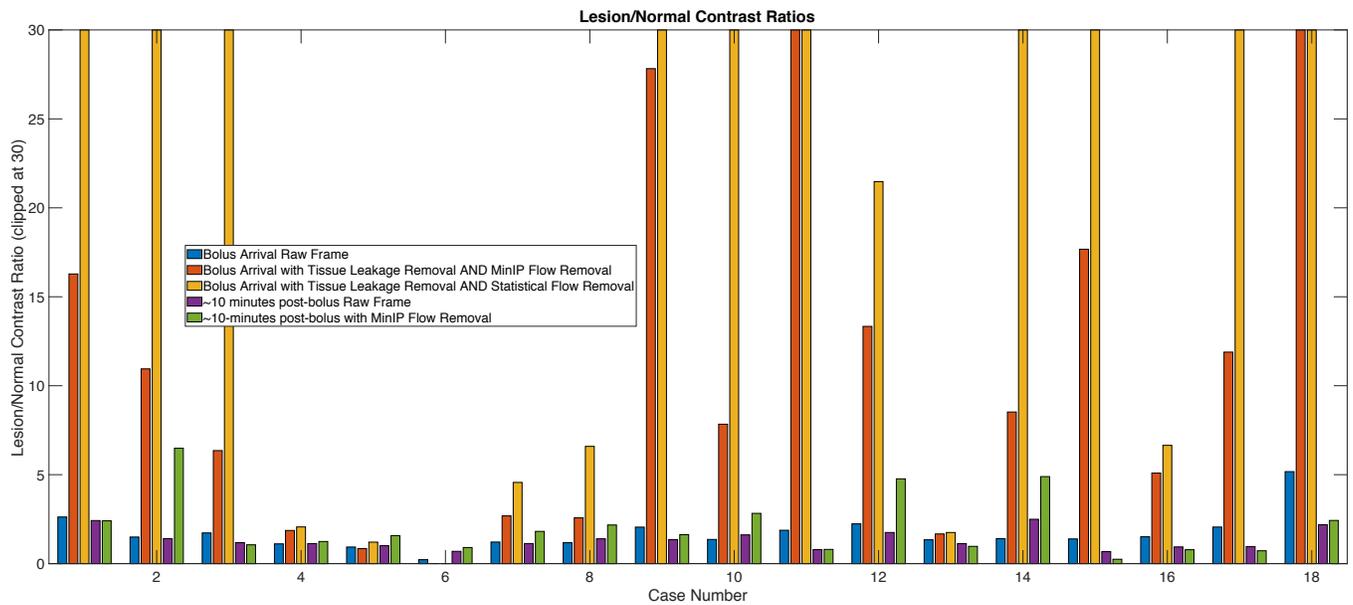

Figure 9: Comparison of Lesion/Normal contrast ratios. Imaging during bolus arrival, with appropriate processing, provides vastly improved ability to distinguish lesions from normal tissue (Gold bars), relative to imaging later after the flow of contrast agent has largely subsided (Purple bars). Case 6 is a false positive (benign lesion) that the Flow&Tissue Leakage disambiguation software correctly eliminated. For the statistical flow removal approach, $\alpha = 2.7$ was used for all cases.

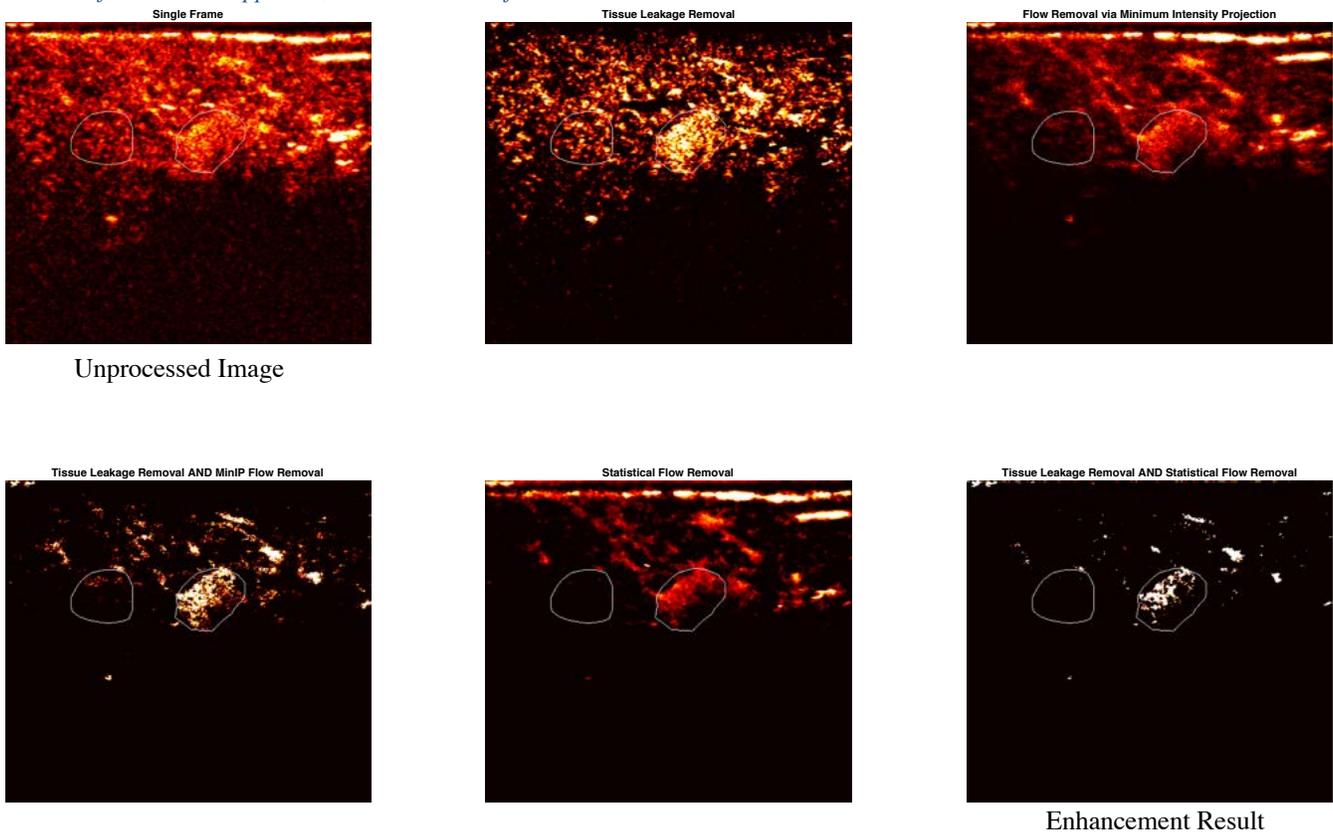

Figure 10: Typical Computational Enhancement Results. Top left: Raw contrast image frame. Top center: Image frame with multi-frame fusion tissue leakage removal applied. Top right: Image derived from minimum-intensity-projection based reflection removal with window size W=20. Bottom left: Image with BOTH minimum-intensity flow removal and tissue leakage removal applied. Note how the two optimizations in combination provide a dramatic improvement. Bottom center: Image frame with statistical flow removal applied, using $\alpha = 2.7$. Bottom right: Image frame with BOTH statistical flow removal and multi-frame fusion tissue leakage removal applied. Comparing the bottom left and bottom-right images illustrates that the statistical flow removal method removes imaging artifacts related to high flow rates that the minimum intensity projection approach is unable to remove.

The outlined regions indicate pathology-proven lesion (right outlined region) vs. normal tissue (left outlined region).

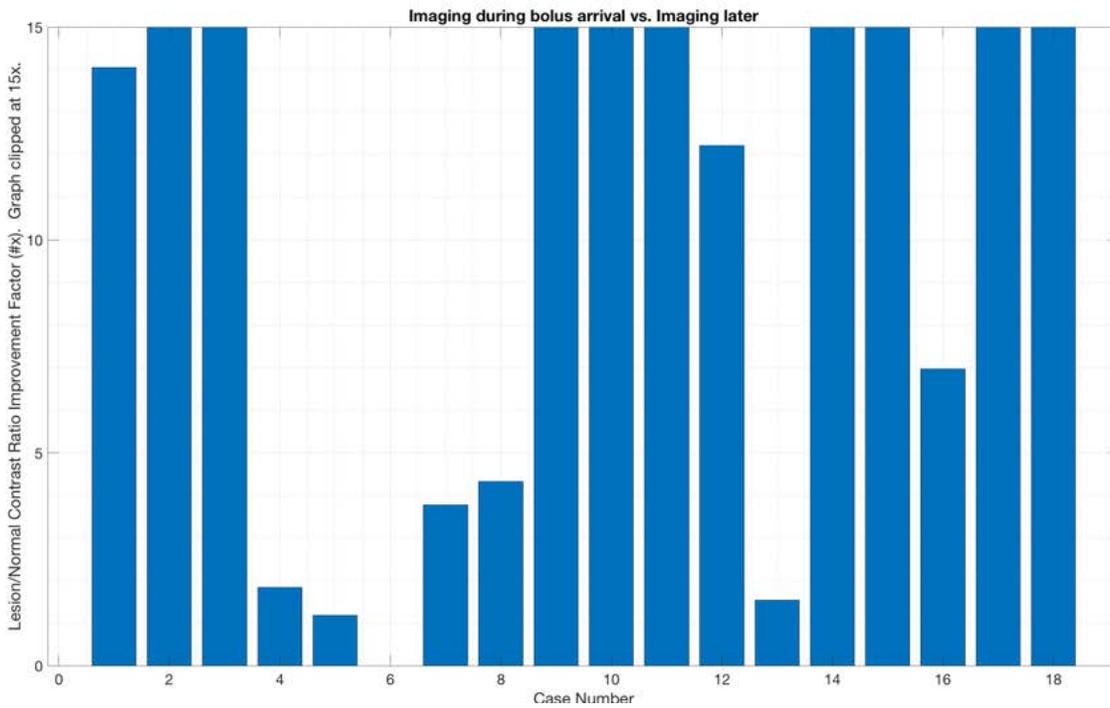

Figure 11: Imaging via computational enhancement during contrast bolus arrival yields substantial improvement in lesion/normal contrast ratio (a key measure of tumor detectability) in comparison to imaging ~11 minutes later, after flow of contrast agent has subsided. Using the statistical method with $\alpha = 2.7$, the contrast ratio improves in all cases, often by an order of magnitude or more. For case #6, the contrast ratio in the computationally-enhanced result drops to 0, eliminating a false positive result in the 11-minute data associated with background tissue leakage that falsely appeared to be a tumor.

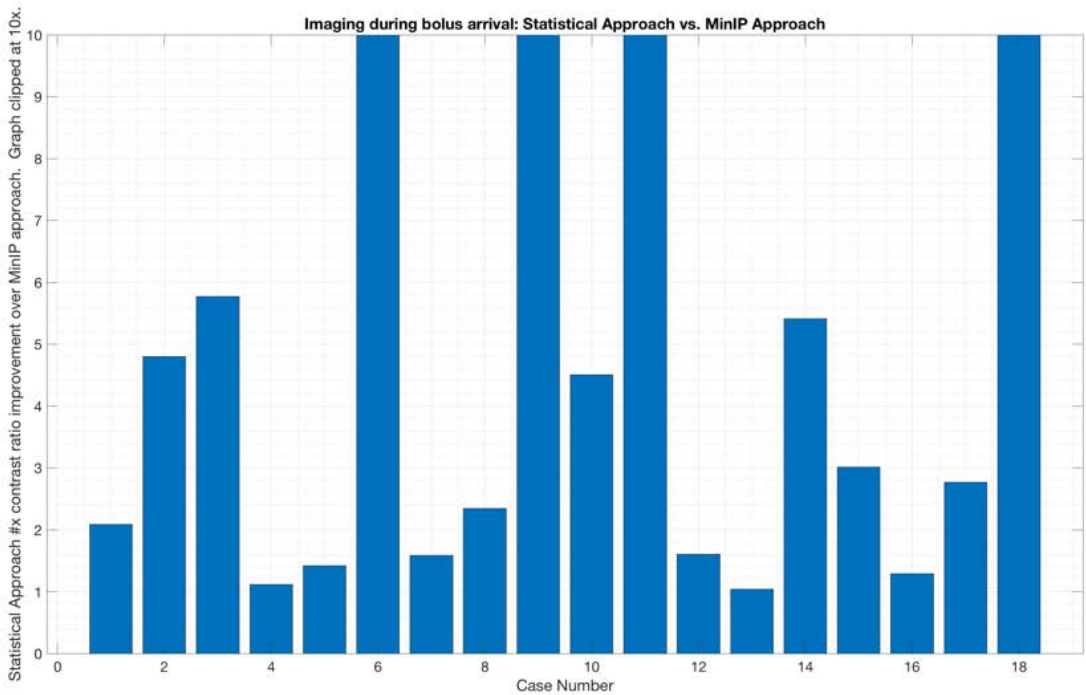

Figure 12: For high flow situations, such as occurs during contrast bolus arrival, the Statistical Approach ($\alpha = 2.7$) provides a substantial improvement in lesion/normal contrast ratio relative to the Minimum Intensity Projection (MinIP) approach.

Even after flow signal removal and tissue leakage signal removal, occasional minor artifacts remain in the images, whose removal should be addressed by future work. Primary among these is accumulation of contrast agent for reasons other than molecularly specific binding.

For future improvements, we are developing cross-window optimizations that apply constraints based on accumulation rate to further distinguish specific binding from other sources of contrast agent accumulation. For example, a large blood vessel filled with contrast agent will arrive quite suddenly in the field of view, filling with contrast agent and saturating the ultrasound detectors far more rapidly than does accumulation due to specific binding. Applying rate-based modeling to computational enhancement results has potential to enable models along the lines of the innovative work of Turco [5] [11] to be applied to data from which the flowing contrast agent signal has been removed. This minimizes the impact of contrast agent flow dynamics (for instance multiple feeding blood vessels) and contrast bolus injection rate on the accumulation rate estimates.

In conclusion, the combination of tissue leakage elimination and statistics-based flow disambiguation provides an order of magnitude improvement in the ability to detect, characterize, and delineate regions of contrast agent accumulation relative to the analysis after flow has subsided approach used in the Willmann 2017 first-in-human study of BR-55 breast imaging.